\documentclass[12pt]{article}
\begin{document}
\setcounter{page}{1}
\def\theequation{\arabic{section}.\arabic{equation}}
\def\theequation{\thesection.\arabic{equation}}
\setcounter{section}{0}

\title{On the reactions p + p $\to$ p + $\Lambda$ + K$^+$ and p + p
$\to$ p + $\Sigma^0$ + K$^+$ near thresholds}

\author{A. Ya. Berdnikov, Ya. A. Berdnikov\,\thanks{E--mail:
berdnikov@twonet.stu.neva.ru} , A. N. Ivanov, \\ V. F. Kosmach,\\
M. D. Scadron\,\thanks{E--mail: scadron@physics.arizona.edu, Physics
Department, University of Arizona, Tucson, Arizona 85721, USA} , and
N. I. Troitskaya}

\date{\today}

\maketitle

\begin{center}
{\it State Technical University of St. Petersburg, Department of
Nuclear Physics, \\ Polytechnicheskaya 29, 195251 St. Petersburg, Russian
Federation}
\end{center}

\begin{center}
\begin{abstract}
The cross sections for the reactions of the strange production p + p
$\to$ p + $\Lambda$ + K$^+$ and p + p $\to$ p + $\Sigma^0$ + K$^+$
near thresholds of the final states p$\Lambda$K$^+$ and
p$\Sigma^0$K$^+$ are calculated in the effective Lagrangian
approach. Our approach is based on the dominant contribution of the
one--pion exchange and strong interaction of the colliding protons in
the initial state. The theoretical values of the cross sections agree
reasonably well with the experimental data. The polarization
properties of the $\Lambda$ and $\Sigma^0$ hyperons are discussed.
\end{abstract}
\end{center}

\newpage

\section{Introduction}
\setcounter{equation}{0}

\hspace{0.2in} Recent experimental data [1--4] on the production of
strangeness in pp collisions, p + p $\to$ p + $\Lambda$ + K$^+$ [1--4]
and p + p $\to$ p + $\Sigma^0$ + K$^+$ [4], for energies of colliding
protons in the region near thresholds of the final states
p$\Lambda$K$^+$ and p$\Sigma^0$K$^+$ have represented experimental
values of the cross sections $\sigma^{\rm pp \to p\Lambda
K^+}(\varepsilon)$ and $\sigma^{\rm pp \to p\Sigma^0
K^+}(\varepsilon)$ with high precision, where $\varepsilon$ is an
excess of energy that we define below. As has been obtained in Ref.[4]
the cross section for the $\Sigma^0$ hyperon, p + p $\to$ p +
$\Sigma^0$ + K$^+$, exceeds by a factor 28 the cross section for the
production of $\Lambda$ hyperon, p + p $\to$ p + $\Lambda$ + K$^+$,
measured for the equivalent excess energies. These data give a
possibility for testing of various theoretical approaches to
mechanisms of a strangeness production from nucleons that is important
for correct description of a strangeness production in heavy--ion
collisions.

The parer is organized as follows. In Section\,1 we calculate the
effective Lagrangians of the transitions p + p $\to$ p + $\Lambda $ +
K$^+$ and p + p $\to$ p + $\Sigma^0$ + K$^+$ in the one--pion exchange
approximation and at leading order in momentum expansion in powers of
the momenta of final state particles. In Section\,2 we calculate the
cross sections for the reactions p + p $\to$ p + $\Lambda $ + K$^+$
and p + p $\to$ p + $\Sigma^0$ + K$^+$ and compare the theoretical
results with the experimental data. In the Conclusion we discuss the
obtained results and the polarization properties of the $\Lambda$ and
$\Sigma^0$ hyperons.

\section{Effective Lagrangians of transitions p + p $\to$ p + 
$\Lambda (\Sigma^0)$ + K$^+$} 
\setcounter{equation}{0}

\hspace{0.2in} In our approach to the description of the reactions p +
p $\to$ p + $\Lambda$ + K$^+$ and p + p $\to$ p + $\Sigma^0$ + K$^+$,
first, we suggest to investigate the transitions p + p $\to$ p +
$\Lambda$ + K$^+$ and p + p $\to$ p + $\Sigma^0$ + K$^+$, where the
wave functions of the protons in the initial pp state are described by
plane waves and all particles in the final states p$\Lambda$K$^+$ and
p$\Sigma^0$K$^+$ are decoupled. These transitions we define by the
effective Lagrangians ${\cal L}^{\rm pp\to p\Lambda K^+}(x)$ and
${\cal L}^{\rm pp\to p\Sigma^0 K^+}(x)$. For the evaluation of these
effective Lagrangian we suggest to use a simplest one--pion exchange
approximation. The Feynman diagrams defining the effective Lagrangians
${\cal L}^{\rm pp\to p\Lambda K^+}(x)$ and ${\cal L}^{\rm pp\to
p\Sigma^0 K^+}(x)$ are depicted in Fig.1.

The analytical expressions corresponding to these diagrams read
\begin{eqnarray}\label{label2.1}
\hspace{-0.5in}&&M(\rm pp\to p\Lambda K^+) =\nonumber\\
\hspace{-0.5in}&&=[\bar{u}(\vec{p}_{\rm p},\alpha_{\rm p})i\gamma^5u(\vec{p}_1,\alpha_1)]
\,\frac{\displaystyle g^2_{\pi NN}}{\displaystyle M^2_{\pi} - (p_{\rm p} -
p_1)^2}\,[\bar{u}(\vec{p}_{\Lambda},\alpha_{\Lambda})i\gamma^5\,
\frac{\displaystyle g_{p\Lambda K^+}}{\displaystyle M_{\rm p} -
\hat{p}_{\Lambda} - \hat{p}_{K}}\,i\gamma^5
u(\vec{p}_2,\alpha_2)]\nonumber\\
\hspace{-0.5in}&& - [\bar{u}(\vec{p}_{\rm p},\alpha_{\rm p})i\gamma^5
u(\vec{p}_2,\alpha_2)]\,\frac{\displaystyle g^2_{\pi
NN}}{\displaystyle M^2_{\pi} - (p_{\rm p} -
p_2)^2}\,[\bar{u}(\vec{p}_{\Lambda},\alpha_{\Lambda})i\gamma^5\,
\frac{\displaystyle g_{p\Lambda K^+}}{\displaystyle M_{\rm p} -
\hat{p}_{\Lambda} - \hat{p}_{K}}\,i\gamma^5 u(\vec{p}_1,\alpha_1)],
\end{eqnarray}  
where $g_{\rm \pi NN} =13.4$ is the coupling constant of the
$\pi$NN interaction, $u(\vec{p}_i,\alpha_i)$ are bispinors of the
protons for $i = 1,2,3$ and the $\Lambda$--hyperon for $i = \Lambda$
with polarizations $\alpha_i$. Then, $M_{\pi} =135\,{\rm MeV}$ and
$M_{\rm p} = 938.3\,{\rm MeV}$ are the masses of a $\pi^0$ meson and a
proton. The amplitude of the transition p + p $\to$ p + $\Sigma^0$ +
K$^+$ can obtained from Eq.(\ref{label2.1}) by a replacement
$g_{p\Lambda K^+} \to g_{p\Sigma^0 K^+}$ and $p_{\Lambda} \to
p_{\Sigma^0}$. We would like to accentuate that we are using the
pseudoscalar couplings for the description of the $\pi^0$ and the
K$^+$ meson coupled to baryons that always fit data well [5].

In the center of mass frames of the colliding protons and the
p$\Lambda$ system the amplitude Eq.(\ref{label2.1}) takes the form
\begin{eqnarray}\label{label2.2}
\hspace{-0.5in}&&M(\rm pp\to p\Lambda K^+) =[\bar{u}(-\vec{q}_{\rm p\Lambda}
-\vec{p}_{\rm K}/2,\alpha_{\rm p})i\gamma^5u(\vec{p},\alpha_1)]\nonumber\\
\hspace{-0.5in}&&\times\,\frac{\displaystyle g^2_{\pi
NN}}{\displaystyle M^2_{\pi} - \Big(\sqrt{M^2_{\rm p} + (\vec{q}_{\rm
p\Lambda} + \vec{p}_{\rm K}/2)^2} - \sqrt{M^2_{\rm p} +\vec{p}^{\;2}}\,\Big)^2
+(\vec{p} + \vec{q}_{\rm p\Lambda} + \vec{p}_{\rm K}/2)^2}\nonumber\\
\hspace{-0.5in}&&\times\,[\bar{u}(\vec{q}_{\rm p\Lambda}
-\vec{p}_{\rm K}/2,\alpha_{\Lambda})i\gamma^5\,\nonumber\\
\hspace{-0.5in}&&\times\,\frac{\displaystyle
g_{p\Lambda K^+}}{\displaystyle M_{\rm p} -
\gamma^0\Big(\sqrt{M^2_{\Lambda} + (\vec{q}_{\rm p\Lambda}
-\vec{p}_{\rm K}/2)^2} + \sqrt{M^2_{\rm K} + \vec{p}^{\;2}_{\rm K}}\,\Big) +
\vec{\gamma}\cdot(\vec{q}_{\rm p\Lambda} +\vec{p}_{\rm K}/2)}\,i\gamma^5
u(-\vec{p},\alpha_2)]\nonumber\\
\hspace{-0.5in}&& -[\bar{u}(-\vec{q}_{\rm p\Lambda}
-\vec{p}_{\rm K}/2,\alpha_{\rm p})i\gamma^5u(- \vec{p},\alpha_2)]\nonumber\\
\hspace{-0.5in}&&\times\,\frac{\displaystyle g^2_{\pi
NN}}{\displaystyle M^2_{\pi} - \Big(\sqrt{M^2_{\rm p} + (\vec{q}_{\rm
p\Lambda} + \vec{p}_{\rm K}/2)^2} - \sqrt{M^2_{\rm p}
+\vec{p}^{\;2}}\,\Big)^2 +(\vec{p} - \vec{q}_{\rm p\Lambda} -
\vec{p}_{\rm K}/2)^2}\nonumber\\
\hspace{-0.5in}&&\times\,[\bar{u}(\vec{q}_{\rm p\Lambda}
-\vec{p}_{\rm K}/2,\alpha_{\Lambda})i\gamma^5\,\nonumber\\
\hspace{-0.5in}&&\times\,\frac{\displaystyle g_{p\Lambda
K^+}}{\displaystyle M_{\rm p} - \gamma^0\Big(\sqrt{M^2_{\Lambda} +
(\vec{q}_{\rm p\Lambda} -\vec{p}_{\rm K}/2)^2} + \sqrt{M^2_{\rm K} +
\vec{p}^{\;2}_{\rm K}}\,\Big) + \vec{\gamma}\cdot(\vec{q}_{\rm
p\Lambda} +\vec{p}_{\rm K}/2)}\,i\gamma^5 u(\vec{p},\alpha_1)],
\end{eqnarray}  
where $\vec{p}_1 = -\vec{p}_2 = \vec{p}\,$ is a relative momentum of
the colliding protons, $\vec{q}_{\rm p\Lambda}$ is a relative momentum
of the p$\Lambda$ system, and $\vec{p}_{\rm K}$ is the momentum of the
K$^+$ meson.

The reaction p + p $\to$ p + $\Lambda (\Sigma^0)$ + K$^+$ is
determined experimentally very close to threshold of the final state
p$\Lambda$K$^+$ (or p$\Sigma^0$K$^+$). The minimum relative
3--momentum of the initial protons is equal to $|\vec{p}\,|_{\rm
threshold}= p_0 =\sqrt{(M_{\Lambda} + M_{\rm K^+} - M_{\rm
p})(M_{\Lambda} + M_{\rm K^+} + 3 M_{\rm p})}/2 = 861.6\,{\rm MeV}$,
where we have used $M_{\Lambda} = 1115.7\,{\rm MeV}$ and $M_{\rm K^+}
= 493.7\,{\rm MeV}$, the masses of the $\Lambda$ hyperon and the K$^+$
meson [6]\footnote{For the reaction p + p $\to$ p + $\Sigma^0$ + K$^+$
the minimum relative momentum of the colliding protons amounts to
$|\vec{p}\,|_{\rm threshold} = p_0 = \sqrt{(M_{\Sigma^0} + M_{\rm K^+}
- M_{\rm p})(M_{\Sigma^0} + M_{\rm K^+} + 3 M_{\rm p})}/2 =
917.5\,{\rm MeV}$ at $M_{\Sigma^0} = 1192.6\,{\rm MeV}$ [6].}. Due to
this close vicinity to threshold the momentum of the K$^+$ meson and
the relative movement of the p$\Lambda$ system (or p$\Sigma^0$) are
smaller compared with all energy scales of the coupled particles. This
allows to expand the matrix element Eq.(\ref{label2.2}) in powers of
$\vec{p}_{\rm K}$ and $\vec{q}_{\rm p\Lambda}$ by keep leading
contributions:
\begin{eqnarray}\label{label2.3}
\hspace{-0.5in}&&M(\rm pp\to p\Lambda K^+) = - \frac{\displaystyle
g_{p\Lambda K^+} g^2_{\pi NN}}{\displaystyle M_{\rm p} + M_{\Lambda} +
M_{\rm K^+}}\,\frac{1}{\displaystyle M^2_{\pi} + 2M_{\rm
p}(\sqrt{M^2_{\rm p} + p^2_0} - M_{\rm p})}\nonumber\\
\hspace{-0.5in}&&\times\,\{[\bar{u}(-\vec{q}_{\rm p\Lambda} -\vec{p}_{\rm
K}/2,\alpha_{\rm p})i\gamma^5u(\vec{p},\alpha_1)]\,[\bar{u}(\vec{q}_{\rm
p\Lambda} -\vec{p}_{\rm K}/2,\alpha_{\Lambda}) u(-\vec{p},\alpha_2)]\nonumber\\
\hspace{-0.5in}&& - \bar{u}(-\vec{q}_{\rm p\Lambda} -\vec{p}_{\rm
K}/2,\alpha_{\rm p})i\gamma^5u(- \vec{p},\alpha_2)]\,[\bar{u}(\vec{q}_{\rm
p\Lambda} -\vec{p}_{\rm K}/2,\alpha_{\Lambda}) u(\vec{p},\alpha_1)]\}.
\end{eqnarray}  
By introducing the effective coupling constant $C_{\rm p\Lambda K^+}$
\begin{eqnarray}\label{label2.4}
C_{\rm p\Lambda K^+} = \frac{\displaystyle g_{p\Lambda K^+} g^2_{\pi
NN}}{\displaystyle M_{\rm p} + M_{\Lambda} + M_{\rm K^+}
}\,\frac{1}{\displaystyle M^2_{\pi} + 2M_{\rm p}(\sqrt{M^2_{\rm p} +
p^2_0} - M_{\rm p})}
\end{eqnarray}  
we can write down the effective Lagrangian ${\cal L}^{\rm pp\to
p\Lambda K^+}(x)$ of the transition p + p $\to$ p + $\Lambda$ +
K$^+$. That reads
\begin{eqnarray}\label{label2.5}
{\cal L}^{\rm pp\to p\Lambda K^+}(x) = -\,C_{\rm p\Lambda
K^+}\,\varphi^{\dagger}_{\rm
K^+}(x)\,[\bar{p}(x)i\gamma^5p(x)][\bar{\Lambda}(x)p(x)],
\end{eqnarray}  
where $p(x)$ and $\Lambda(x)$ are the operators of the interpolating
proton and $\Lambda$--hyperon fields, and $\varphi^{\dagger}_{\rm
K^+}(x)$ is the operator of the interpolating K$^+$--meson field. By
making the replacement $\Lambda \to \Sigma^0$ in Eq.(\ref{label2.5})
we obtain the effective Lagrangian ${\cal L}^{\rm pp\to p\Sigma^0
K^+}(x)$ of the transition p + p $\to$ p + $\Sigma^0$ + K$^+$.

It is convenient to represent the effective Lagrangian in terms of the
interactions describing the p$\Lambda$ system in the certain spin
state. This can be carried out by means of a Fierz transformation [7].
By performing a Fierz transformation we recast the effective
Lagrangian ${\cal L}^{\rm pp\to p\Lambda K^+}(x)$ into the form
\begin{eqnarray}\label{label2.6}
{\cal L}^{\rm pp\to p\Lambda K^+}(x) &=& i\,\frac{1}{4}\,C_{\rm
p\Lambda K^+}\,\varphi^{\dagger}_{\rm K^+}(x)\,
\{[\bar{p}(x)\gamma^5\Lambda^c(x)][\bar{p^c}(x)p(x)]\nonumber\\ && +
[\bar{p}(x)\Lambda^c(x)][\bar{p^c}(x)\gamma^5p(x)] +
[\bar{p}(x)\gamma^{\mu}\Lambda^c(x)][\bar{p^c}(x)\gamma_{\mu}\gamma^5p(x)]\}.
\end{eqnarray}  
The first term and the last two in the Lagrangian Eq.(\ref{label2.6})
describe the p$\Lambda$ system coupled in the spin singlet and triplet
state, respectively.

Since near threshold the p$\Lambda$ (or p$\Sigma^0$) system couples
mainly in the spin singlet state, ${^1}{\rm S}_0$, we should leave
only the first term. This yields the following effective Lagrangian
\begin{eqnarray}\label{label2.7}
{\cal L}^{\rm pp\to p\Lambda K^+}(x)= i\,\frac{1}{4}\,C_{\rm p\Lambda
K^+}\,\varphi^{\dagger}_{\rm K^+}(x)\,
[\bar{p}(x)\gamma^5\Lambda^c(x)][\bar{p^c}(x)p(x)].
\end{eqnarray}  
The wave functions of the particles in the transition p + p $\to$ p +
$\Lambda$ + K$^+$ are plane waves. In order to describe a physical
reaction p + p $\to$ p + $\Lambda$ + K$^+$ we suggest to take into
account interactions between particles both in the final and in the
initial state.

\section{Cross sections for near threshold reactions p + p $\to$ p + 
$\Lambda(\Sigma^0)$ + K$^+$} 
\setcounter{equation}{0}

\hspace{0.2in} The contribution of the interaction in the
p$\Lambda$--system can be obtained by summing up an infinite series of
one--baryon loop diagrams with a point--like
$(\bar{p}\bar{\Lambda})(p\Lambda)$ coupling describing a low--energy
transition p + $\Lambda$ $\to$ p + $\Lambda$ [7].  After the
evaluation of momentum integrals and the renormalization of the wave
functions of the proton and the $\Lambda$ hyperon [7] we can represent
the contribution of this series in the phenomenological form in terms
of the S--wave scattering length $a_{\rm p\Lambda}$ and the effective
range $r_{\rm p\Lambda}$:
\begin{eqnarray}\label{label3.1}
f^{\rm p\Lambda \to p\Lambda}(q_{\rm p\Lambda}) =
\frac{1}{\displaystyle 1 - \frac{1}{2}\,a_{\rm p\Lambda}r_{\rm
p\Lambda}q^2_{\rm p\Lambda} + i\,a_{\rm p\Lambda}q_{\rm p\Lambda}}.
\end{eqnarray}  
This is well--known Watson form for the final--state interaction [8]
having been used by Balewski {\it et al.} [9] for the description of
the final p$\Lambda$ interaction in the reaction p + p $\to$ p +
$\Lambda$ + K$^+$. Below we use the numerical values of the S--wave
scattering length and the effective range, $ a_{\rm p\Lambda} =
-2.0\,{\rm fm}$ and $r_{\rm p\Lambda} = 1.0\,{\rm fm}$, recommended by
Balewski {\it et al.} [9]. We would like to emphasize that since
finally the contribution of the final--state interaction is expressed
in terms of phenomenological parameters, the S--wave scattering length
and the effective range taken from experimental data, a knowledge of
an explicit value of a coupling constant of a local
$(\bar{p}\bar{\Lambda})(p\Lambda)$ interaction describing a
low--energy transition p + $\Lambda$ $\to$ p + $\Lambda$ and defining
vertices in the one--baryon loop diagrams is not important [7]. For
the analysis of elastic low--energy p$\Sigma^0$ scattering we assume
that $a_{\rm p\Sigma^0} = a_{\rm p\Lambda} = -2.0\,{\rm fm}$ and
$r_{\rm p\Sigma^0} = r_{\rm p\Lambda} = 1.0\,{\rm fm}$. Below we show
that this assumption does not contradict the experimental data [4].

Unlike the p$\Lambda$ interaction in the final state in order to
describe the interaction in the initial pp state we have to specify
the coupling constant of the transition p + p $\to$ p + p. Since the
relative momentum of the pp state is comeasurable with a mass of a
proton, so that a Coulomb repulsion between protons can be
neglected. We suggest to describe the pp interaction in the one--pion
exchange approximation. As experimentally relative momenta of the pp
state differ slightly from the threshold momentum, we can represent
the pp interaction describing the transition p + p $\to$ p + p in the
following local form
\begin{eqnarray}\label{label3.2}
{\cal L}^{\rm pp\to pp}(x) = \frac{1}{8}\,C_{\rm
pp}\,[\bar{p}(x)p^c(x)][\bar{p^c}(x)p(x)].
\end{eqnarray}  
This effective Lagrangian describes the pp system coupled in the
spin--triplet state. In the factor $1/8$ the multiplier $1/4$ is
caused by a Fierz transformation of the one--pion exchange interaction
[7]. The phenomenological coupling constant $C_{\rm pp}$ is then
defined by
\begin{eqnarray}\label{label3.3}
C_{\rm pp} = \frac{g^2_{\rm \pi NN}}{4\vec{p}^{\,2}}\,{\ell n}\Bigg(1 +
\frac{4\vec{p}^{\,2}}{M^2_{\pi}}\Bigg).
\end{eqnarray}  
It is obtained from the one--pion exchange diagram of the transition p
+ p $\to$ p + p by averaging over all possible directions of a
relative momentum of the final pp state. We would like to accentuate
that since relative momenta of the incident protons are very close to
threshold, the quantity $C_{\rm pp}$ is practically constant
determined by $\vec{p}^{\,2} \simeq p^2_0$ at $p_0 = 861.4\,{\rm MeV}$
or $p_0 = 917.5\,{\rm MeV}$ for the reactions p + p $\to$ p $\Lambda$
+ K$^+$ and p + p $\to$ p $\Sigma^0$ + K$^+$, respectively.

By summing up an infinite series of one--proton loop diagrams the
vertices of which are defined by the effective interaction
Eq.(\ref{label3.2}) we arrive at the expression [7]
\begin{eqnarray}\label{label3.4}
[\bar{u^c}(-\vec{p},\alpha_2)u(\vec{p},\alpha_1)] \to
\frac{[\bar{u^c}(-\vec{p},\alpha_2)u(\vec{p},\alpha_1)]}{\displaystyle
1 + \frac{C_{\rm pp}}{64\pi^2}\int\frac{d^4k}{\pi^2 i}\,{\rm
tr}\Bigg\{\frac{1}{M_{\rm p} - \hat{k}}\frac{1}{M_{\rm p} - \hat{k} -
\hat{P}}\Bigg\}},
\end{eqnarray}  
where $P = (2\,\sqrt{\vec{p}^{\,2} + M^2_{\rm p}},\vec{0}\,)$.

After the subtraction of trivial $\vec{p}$--independent divergent
contributions and the renormalization of the wave functions of the
protons we obtain the contribution of the interaction of the protons
in the initial state
\begin{eqnarray}\label{label3.5}
&&[\bar{u^c}(-\vec{p},\alpha_2)u(\vec{p},\alpha_1)] \to
\frac{[\bar{u^c}(-\vec{p},\alpha_2)u(\vec{p},\alpha_1)]}{\displaystyle
1 + \frac{C_{\rm pp}}{64\pi^2}\int\frac{d^4k}{\pi^2 i}\,{\rm
tr}\Bigg\{\frac{1}{M_{\rm p} - \hat{k}}\frac{1}{M_{\rm p} - \hat{k} -
\hat{P}}\Bigg\}}\nonumber\\ &&\to
\frac{[\bar{u^c}(-\vec{p},\alpha_2)u(\vec{p},\alpha_1)]}{\displaystyle
1 + \frac{C_{\rm
pp}(\vec{p}^{\,2},\Lambda)}{8\pi^2}\,\frac{|\vec{p}\,|^3}{\sqrt{\vec{p}^{\,2}
+ M^2_{\rm p}}}\left[{\ell n}\left(\frac{\sqrt{\vec{p}^{\,2} +
M^2_{\rm p}} + |\vec{p}\,|}{\sqrt{\vec{p}^{\,2} + M^2_{\rm p}} -
|\vec{p}\,|}\right) +\pi\,i\,\right]},
\end{eqnarray}  
where $C_{\rm pp}(\vec{p}^{\,2},\Lambda)$ is given by
\begin{eqnarray}\label{label3.6}
C_{\rm pp}(\vec{p}^{\,2},\Lambda) = \frac{\displaystyle C_{\rm
pp}}{\displaystyle 1 + \frac{C_{\rm
pp}\vec{p}^{\,2}}{4\pi^2}\left[{\ell n}\left(\frac{\Lambda}{M_{\rm N}}
+\sqrt{1 + \frac{\Lambda^2}{M^2_{\rm N}}}\right) -
\frac{\Lambda}{\sqrt{M^2_{\rm N} + \Lambda^2}}\right]}.
\end{eqnarray}  
The appearance of the cut--off $\Lambda$ is caused by non--trivial
$\vec{p}$--dependent logarithmically divergent contributions.  The
cut--off $\Lambda$ restricts from above 3--momenta of virtual proton
fluctuations. Since our approach is an effective one, so the
dependence of the amplitude on the cut--off seems to be usual
[10,11]. The only thing one needs is to choose the value of the
cut--off by an appropriate physical way. In our case it is reasonable
to have $\Lambda$ to be of order a mass of a resonance nearest to a
nucleon, that is the $\Delta(1232)$ resonance [6]. Therefore, in our
calculations we would set $\Lambda = 1200\,{\rm MeV}$.

Thus, the amplitude of the reaction p + p $\to$ p + $\Lambda$ + K$^+$
near threshold of the final p$\Lambda$K$^+$ state is defined by
\begin{eqnarray}\label{label3.7}
\hspace{-0.5in}&&{\cal M}({\rm pp \to p\Lambda K^+}) =
\frac{1}{2}\,C_{\rm p\Lambda K^+}\,\frac{[\bar{u}(-\vec{q}_{\rm
p\Lambda} -\vec{p}_{\rm K}/2,\alpha_{\rm
p})i\gamma^5 u^c(\vec{q}_{\rm p\Lambda} -\vec{p}_{\rm
K}/2,\alpha_{\Lambda})]}{\displaystyle 1 - \frac{1}{2}\,a_{\rm
p\Lambda}r_{\rm p\Lambda}q^2_{\rm p\Lambda} + i\,a_{\rm
p\Lambda}q_{\rm p\Lambda}}\nonumber\\
\hspace{-0.5in}&&\times\,
\frac{[\bar{u^c}(-\vec{p},\alpha_2)u(\vec{p},\alpha_1)]}
{\displaystyle 1 + \frac{C_{\rm
pp}(\vec{p}^{\,2},\Lambda)}{8\pi^2}\,\frac{|\vec{p}\,|^3}{\sqrt{\vec{p}^{\,2}
+ M^2_{\rm p}}}\left[{\ell n}\left(\frac{\sqrt{\vec{p}^{\,2} +
M^2_{\rm p}} + |\vec{p}\,|}{\sqrt{\vec{p}^{\,2} + M^2_{\rm p}} -
|\vec{p}\,|}\right) +\pi\,i\,\right]}\nonumber\\
\hspace{-0.5in}&&\times\,\sqrt{ \frac{M_{\rm
pK^+}}{q_{\rm pK^+}}\,\frac{2\pi \alpha }{\displaystyle e^{\textstyle
2\pi \alpha M_{\rm pK^+}/q_{\rm pK^+}} - 1}},
\end{eqnarray}  
where the last factor depending of $\alpha =1/137$, the fine structure
constant, and $q_{\rm pK^+}$, a relative momentum of the pK$^+$
system, takes into account the Coulomb repulsion between the daughter
proton and the K$^+$ meson at low relative energies [9] ( see also
[7]), $M_{\rm pK^+} = M_{\rm p}M_{\rm K^+}/(M_{\rm p} + M_{\rm K^+})$
is the reduced mass of the pK$^+$ system.

Then, relative momenta of the pp system are very close to threshold,
$|\vec{p}\,|\simeq p_0 = 861.4\,{\rm MeV}$. Hence, we can calculate
the contribution of the interactions in the pp state numerically. This
gives
\begin{eqnarray}\label{label3.8}
\frac{1} {\displaystyle 1 + \frac{C_{\rm
pp}(p^2_0,\Lambda)}{8\pi^2}\,\frac{p^3_0}{\sqrt{p^2_0 + M^2_{\rm
p}}}\left[{\ell n}\left(\frac{\sqrt{p^2_0 + M^2_{\rm p}} +
p_0}{\sqrt{p^2_0 + M^2_{\rm p}} - p_0}\right) +\pi\,i\,\right]} =
0.308\,e^{\textstyle -i\,46.6^0}.
\end{eqnarray}  
For the reaction p + p $\to$ p + $\Sigma^0$ + K$^+$ we get $0.294\,
e^{\textstyle -i\,46.3^0}$.

By calculating numerically a part of the coupling constant $C_{\rm
p\Lambda K^+}$ given by Eq.(\ref{label2.4}) we reduce the amplitude of
the reaction p + p $\to$ p + $\Lambda$ + K$^+$ to the following form
\begin{eqnarray}\label{label3.9}
&&{\cal M}({\rm pp \to p\Lambda K^+}) =
1.676\times 10^{-8}\,e^{\textstyle -i\,46.6^0}\,\frac{\displaystyle g_{\rm
p\Lambda K^+}}{\displaystyle 1 - \frac{1}{2}\,a_{\rm p\Lambda}r_{\rm
p\Lambda}q^2_{\rm p\Lambda} + i\,a_{\rm p\Lambda}q_{\rm
p\Lambda}}\nonumber\\
&&\times\,[\bar{u}(-\vec{q}_{\rm p\Lambda}
-\vec{p}_{\rm K}/2,\alpha_{\rm p})i\gamma^5 u^c(\vec{q}_{\rm p\Lambda}
-\vec{p}_{\rm K}/2,\alpha_{\Lambda})][\bar{u^c}(-\vec{p},\alpha_2)
u(\vec{p},\alpha_1)]\nonumber\\
&&\times\,\sqrt{\frac{M_{\rm pK^+}}{q_{\rm
pK^+}}\,\frac{2\pi \alpha }{\displaystyle e^{\textstyle 2\pi \alpha
M_{\rm pK^+}/q_{\rm pK^+}} - 1}}.
\end{eqnarray}  
For the reaction p + p $\to$ p + $\Sigma^0$ + K$^+$ the numerical
factor is equal to $1.433\times 10^{-8}\,e^{\textstyle -i\,46.3^0}$.  

The amplitude squared, averaged over polarizations of the initial
protons and summed over polarizations of the final baryons amounts to
\begin{eqnarray}\label{label3.10}
&&\overline{|{\cal M}({\rm pp \to p\Lambda K^+})|^2}
=4.494\times10^{-15}\,p^2_0 M_{\rm p}M_{\Lambda}\nonumber\\
&&\times\, \frac{\displaystyle \,g^2_{\rm p\Lambda
K^+}}{\displaystyle\Big( 1 -
\frac{1}{2}\,a_{\rm p\Lambda}r_{\rm p\Lambda}q^2_{\rm p\Lambda}\Big)^2
+ a^2_{\rm p\Lambda}q^2_{\rm p\Lambda}}\,\frac{M_{\rm pK^+}}{q_{\rm
pK^+}}\,\frac{2\pi \alpha }{\displaystyle e^{\textstyle 2\pi \alpha
M_{\rm pK^+}/q_{\rm pK^+}} - 1}.
\end{eqnarray}  
For the reaction p + p $\to$ p + $\Sigma^0$ + K$^+$ the numerical
factor acquires the value $3.285\times10^{-15}$.

The cross section for the reaction p + p $\to$ p + $\Lambda$ + K$^+$
reads
\begin{eqnarray}\label{label3.11}
\sigma_{\rm p\Lambda K^+}(\varepsilon) = 0.043\,g^2_{\rm
p\Lambda K^+}\,\varepsilon^2\,\Omega_{\rm p\Lambda K^+}(\varepsilon),
\end{eqnarray}  
where the cross section and the excess of energy $\varepsilon =
2\,\sqrt{\vec{p}^{\,2} + M^2_{\rm p}} - M_{\rm p} - M_{\Lambda} -
M_{\rm K^+}$ are measured in (nb) and (MeV), respectively. The
function $\Omega_{\rm p\Lambda K^+}(\varepsilon)$ related to the phase
volume of the reaction is defined by
\begin{eqnarray}\label{label3.12}
&&\Omega_{\rm p\Lambda K^+}(\varepsilon) = \frac{1}{4\pi^3
\varepsilon^2}\Bigg(\frac{M_{\rm p} + M_{\Lambda} + M_{\rm
K^+}}{M_{\rm p}M_{\Lambda}M_{\rm K^+}}\Bigg)^{3/2}\nonumber\\
&&\times\int\frac{\displaystyle \delta^{(3)}(\vec{p}_{\rm p} +
\vec{p}_{\Lambda} + \vec{p}_{\rm K})}{\displaystyle\Big( 1 -
\frac{1}{2}\,a_{\rm p\Lambda}r_{\rm p\Lambda}q^2_{\rm p\Lambda}\Big)^2
+ a^2_{\rm p\Lambda}q^2_{\rm p\Lambda}}\,\frac{M_{\rm pK^+}}{q_{\rm
pK^+}}\,\frac{2\pi \alpha }{\displaystyle e^{\textstyle 2\pi \alpha
M_{\rm pK^+}/q_{\rm pK^+}} - 1}\nonumber\\
&&\times\,\delta\Bigg(\varepsilon - \frac{\vec{p}^{\,2}_{\rm
p}}{2M_{\rm p}} - \frac{\vec{p}^{\,2}_{\Lambda}}{2M_{\Lambda}} -
\frac{\vec{p}^{\,2}_{\rm K}}{2M_{\rm K^+}}\Bigg)\,d^3p_{\rm
K}d^3p_{\rm p} d^3p_{\Lambda}
\end{eqnarray}  
and normalized to unity at $\alpha \to 0$ and $a_{\rm p\Lambda} \to
0$.

The cross section for the reaction p + p $\to$ p + $\Sigma^0$ + K$^+$
can be evaluated in analogy with the reaction p + p $\to$ p +
$\Lambda$ + K$^+$ and reads
\begin{eqnarray}\label{label3.13}
\sigma_{\rm p\Sigma^0 K^+}(\varepsilon) = 0.035\,g^2_{\rm
p\Sigma^0 K^+}\,\varepsilon^2\,\Omega_{\rm p\Sigma^0 K^+}(\varepsilon).
\end{eqnarray}  
The function $\Omega_{\rm p\Sigma^0 K^+}(\varepsilon)$ results from
Eq.(\ref{label3.12}) via a replacement $\Lambda \to \Sigma^0$.

In terms of the axial--vector coupling constants $D$ and $F$ and
$g_{\rm \pi NN}$ the coupling constants $g_{\rm p\Lambda K^+}$ and
$g_{\rm p\Sigma^0 K^+}$ are defined by [12]
\begin{eqnarray}\label{label3.14}
g_{\rm p\Lambda K^+} &=& -\frac{1}{\sqrt{3}}\,\Bigg(\frac{D + 3F}{D +
F}\Bigg)\,g_{\rm \pi NN}, \nonumber\\ 
g_{\rm p\Sigma^0 K^+} &=& -
\Bigg(\frac{D - F}{D + F}\Bigg)\,g_{\rm \pi NN}.
\end{eqnarray}  
The cross sections Eqs.(\ref{label3.11}) and (\ref{label3.13}) then
read
\begin{eqnarray}\label{label3.15}
\sigma^{\rm pp\to p\Lambda K^+}(\varepsilon) &=&2.576\,\Bigg(\frac{D +
3F}{D + F}\Bigg)^2\,\varepsilon^2\,\Omega_{\rm p\Lambda
K^+}(\varepsilon),\nonumber\\ \sigma^{\rm pp\to p\Sigma^0
K^+}(\varepsilon) &=& 6.208\,\Bigg(\frac{D - F}{D +
F}\Bigg)^2\,\varepsilon^2\,\Omega_{\rm p\Sigma^0 K^+}(\varepsilon).
\end{eqnarray}  
For the numerical analysis the functions $\Omega_{\rm p\Lambda
K^+}(\varepsilon)$ and $\Omega_{\rm p\Sigma^0 K^+}(\varepsilon)$ can
be given in the more convenient form
\begin{eqnarray}\label{label3.16}
\hspace{-0.5in}&&\Omega_{\rm p\Lambda K^+}(\varepsilon) = \frac{2}{\pi
\varepsilon^2}\frac{M_{\rm p} +
M_{\Lambda}}{M_{\Lambda}}\sqrt{\frac{M_{\rm p}M_{\Lambda}M_{\rm
K^+}}{M_{\rm p} + M_{\Lambda} + M_{\rm K^+}}}\nonumber\\
\hspace{-0.5in}&&\times\int\limits^{\textstyle\varepsilon}_{\textstyle
0}\frac{1}{\displaystyle ( 1 - a_{\rm p\Lambda}r_{\rm p\Lambda}M_{\rm
p\Lambda}T_{\rm p\Lambda})^2 +2 a^2_{\rm p\Lambda}M_{\rm
p\Lambda}T_{\rm p\Lambda}}\int\limits^{\textstyle v^+_{\rm
pK^+}}_{\textstyle v^-_{\rm pK^+}}\frac{2\pi \alpha }{\displaystyle
e^{\textstyle 2\pi \alpha /v_{\rm pK^+}} - 1} dv_{\rm pK^+} dT_{\rm
p\Lambda},
\end{eqnarray}  
where we have denoted
\begin{eqnarray}\label{label3.17}
v^+_{\rm pK^+} &=&\sqrt{\frac{2(M_{\rm p} + M_{\Lambda} + M_{\rm
K^+})}{M_{\rm K^+}(M_{\rm p} + M_{\Lambda})}\,(\varepsilon - T_{\rm
p\Lambda})} + \sqrt{\frac{2M_{\Lambda}}{M_{\rm p}}\frac{T_{\rm
p\Lambda}}{M_{\rm p} + M_{\Lambda}}},\nonumber\\
v^-_{\rm pK^+} &=&\left|\sqrt{\frac{2(M_{\rm p} + M_{\Lambda} + M_{\rm
K^+})}{M_{\rm K^+}(M_{\rm p} + M_{\Lambda})}\,(\varepsilon - T_{\rm
p\Lambda})} - \sqrt{\frac{2M_{\Lambda}}{M_{\rm p}}\frac{T_{\rm
p\Lambda}}{M_{\rm p} + M_{\Lambda}}}\right|.
\end{eqnarray}  
The function $\Omega_{\rm p\Sigma^0 K^+}(\varepsilon)$ should be
obtained from the function $\Omega_{\rm p\Lambda K^+}(\varepsilon)$
given by Eq.(\ref{label3.16}) via a simple replacement $M_{\Lambda}
\to M_{\Sigma^0}$. The numerical values of these functions are
tabulated in Tables 1 and 2, respectively.

For numerical calculations of the cross sections we use $F = 0.459\pm
0.008$ and $D = 0.798 \pm 0.008$ [13]. The numerical values of the
cross sections for the excess of energy $\varepsilon$ ranging over the
region $0.68\,{\rm MeV} \le \varepsilon \le 6.68\,{\rm MeV}$ [1,2,4]
are adduced in Table 1 and 2.

\section{Conclusion}
\setcounter{equation}{0}

\hspace{0.2in} We have developed a phenomenological approach to the
description of the reactions p + p $\to$ p + $\Lambda$ + K$^+$ and p +
p $\to$ p + $\Sigma^0$ + K$^+$ near thresholds of the final states
p$\Lambda$K$^+$ and p$\Sigma^0$K$^+$, respectively. 

The theoretical cross section for the reaction p + p $\to$ p +
$\Lambda$ + K$^+$ agrees reasonably well with the experimental
data. The numerical values of the cross sections are adduced in Table
1. We show that for the excess of energy $\varepsilon$ ranging over
the region $0.68\,{\rm MeV} \le \varepsilon \le 6.68\,{\rm MeV}$ the
cross section is proportional to $\varepsilon^2$, $\sigma^{\rm pp\to
p\Lambda K^+}(\varepsilon) =(4.5\pm 0.1)\,\varepsilon^2\,{\rm nb}$
. This fits well the experimental value $\sigma^{\rm pp\to p\Lambda
K^+}(\varepsilon) =(4.4\pm 0.7)\,\varepsilon^2\,{\rm nb}$ [2].

The cross section for the reaction p + p $\to$ p + $\Lambda$ + K$^+$
has been also measured for the higher excess of energies [4]:
$\sigma^{\rm pp\to p\Lambda K^+}(\varepsilon = 8.6\,{\rm MeV})
=(264\pm 20)\,{\rm nb}$, $\sigma^{\rm pp\to p\Lambda K^+}(\varepsilon
= 10.9\,{\rm MeV}) =(392\pm 33)\,{\rm nb}$ and $\sigma^{\rm pp\to
p\Lambda K^+}(\varepsilon = 13.2\,{\rm MeV}) =(534\pm 47)\,{\rm
nb}$. Our theoretical predictions for these energies read:
$\sigma^{\rm pp\to p\Lambda K^+}(\varepsilon = 8.6\,{\rm MeV})
=(298\pm 6)\,{\rm nb}$, $\sigma^{\rm pp\to p\Lambda K^+}(\varepsilon =
10.9\,{\rm MeV}) =(444\pm 9)\,{\rm nb}$ and $\sigma^{\rm pp\to
p\Lambda K^+}(\varepsilon = 13.2\,{\rm MeV}) =(604\pm 12)\,{\rm
nb}$. For the calculation of these cross sections we have also taken
into account the momentum dependence of the structure function $C_{\rm
pp}(\vec{p}^{\,2}, \Lambda)$. 

We would like to emphasize that within our approach the theoretical
cross section for the reaction p + p $\to$ p + $\Lambda$ + K$^+$ fits
well the experimental values just at $\varepsilon = 138\,{\rm
MeV}$. Really, at $\varepsilon = 138\,{\rm MeV}$ the theoretical value
of the cross section $\sigma^{\rm pp\to p\Lambda K^+}(\varepsilon =
138{\rm MeV})= (13.2\pm 0.3)\,{\rm \mu b}$ agrees well with the
experimental one $\sigma^{\rm pp\to p\Lambda K^+}(\varepsilon =
138\,{\rm MeV}) = (12.0\pm 0.4)\,{\rm \mu b}$ [3]. In average the
accuracy of the agreement between the theoretical cross section for
the reaction p + p $\to$ p + $\Lambda$ + K$^+$ and the experimental
data [2--4] is about 11$\%$.

However, we cannot pass by the fact that the experimental value of the
cross section measured at $\varepsilon = 55\,{\rm MeV}$ [3]:
$\sigma^{\rm pp\to p\Lambda K^+}(\varepsilon = 55\,{\rm MeV}) =(2.7\pm
0.3)\,{\rm \mu b}$ is by a factor 1.7 smaller compared with the
theoretical one: $\sigma^{\rm pp\to p\Lambda K^+}(\varepsilon =
55\,{\rm MeV}) =(4.7\pm 0.3)\,{\rm \mu b}$.  Since the theoretical
cross section in the excess of energy region $0.68\,{\rm MeV} \le
\varepsilon \le 138\,{\rm MeV}$ is a smooth function of $\varepsilon$
proportional to $\varepsilon^2$ and fits reasonably well the
experimental value of the cross section at $\varepsilon = 138\,{\rm
MeV}$, we argue that the result obtained at $\varepsilon = 55\,{\rm
MeV}$ seems to be underestimated and demands to be remeasured.

The cross section for the reaction p + p $\to$ p + $\Sigma^0$ + K$^+$
is also described well in our approach. The theoretical values of the
cross section adduced in Table 2 are in reasonable agreement with the
experimental data for all excess of energies from the interval
$0.68\,{\rm MeV} \le \varepsilon \le 6.68\,{\rm MeV}$. In this energy
region the theoretical cross section is proportional to
$\varepsilon^2$. The average value $\sigma^{\rm pp\to
p\Sigma^0K^+}(\varepsilon)= (0.26\pm 0.03)\,\varepsilon^2\,{\rm nb}$
fits well the experimental data $\sigma^{\rm pp\to
p\Sigma^0K^+}(\varepsilon)= (0.22\pm 0.11)\,\varepsilon^2\,{\rm nb}$
[4]. At $\varepsilon = 138\,{\rm MeV}$ we predict $\sigma^{\rm pp\to
p\Sigma^0K^+}(\varepsilon = 138\,{\rm MeV})= (0.72\pm 0.08)\,{\rm \mu
b}$ that agrees well with the experimental value $\sigma^{\rm pp\to
p\Sigma^0K^+}(\varepsilon = 138\,{\rm MeV})= (1.0\pm 0.5)\,{\rm \mu
b}$ [3].

In our approach the enhancement of the cross section for the reaction
p + p $\to$ p + $\Lambda$ + K$^+$ with respect to the cross section
for the reaction p + p $\to$ p + $\Sigma^0$ + K$^+$ is completely a
unitary symmetry effect. In fact, the coupling constant $g_{\rm
p\Sigma^0 K^+}$ is smaller compared by a factor of 0.16 with the
coupling constant $g_{\rm p\Lambda K^+}$. This is in agreement with
the conclusion given by Kaiser [11]. However, unlike the Kaiser's
approach to the description of the reactions p + p $\to$ p + $\Lambda$
+ K$^+$ and p + p $\to$ p + $\Sigma^0$ + K$^+$ we point out the
dominant role of the contribution of the one--pion exchange and the
strong interaction of the protons in the initial state.

The key--point of our approach to the description of the protons
coupled in the initial state is a reduction of pp interaction to a
local form via a phenomenological interaction Eq.(\ref{label3.3})
based on the one--pion exchange. By having fixed the phenomenological
coupling of a four--proton interaction we have succeeded then in
deriving the contribution of this interaction to the amplitudes of the
reactions p + p $\to$ p + $\Lambda$ + K$^+$ and p + p $\to$ p +
$\Sigma^0$ + K$^+$ via the summation of an infinite series of
one--proton loop diagrams. After the evaluation of these diagrams and
the renormalization of the wave functions of the protons we have
arrived at the expression that has turned out to be dependent on a
cut--off $\Lambda$ restricting from above 3--momenta of virtual proton
fluctuations. The appearance of a dependence on a cut--off is a usual
case in a phenomenological approach to the description of the
reactions under consideration [10,11]. The main point is to fix this
parameter in an appropriate way.  The neglect of the contribution of
baryon resonances to the amplitude of the pp interaction makes a hint
that the cut--off $\Lambda$ should be of order of the mass of a
nearest resonance that is the $\Delta(1232)$ resonance. That is why we
have set $\Lambda = 1200\,{\rm MeV}$. As has been discussed above this
has led to the description of the cross sections for the reactions p +
p $\to$ p + $\Lambda$ + K$^+$ and p + p $\to$ p + $\Sigma^0$ + K$^+$
with accuracy about 11$\%$.

We would like to underscore that our approach to the description of
the protons coupled in the initial state is rather similar
ideologically and technically to that having been applied by Achasov
{\it et al.} [14] to the analysis of the contribution of the scalar
$a_0(980)$ and $f_0(980)$ mesons treated as four--quark states [15] to
the amplitudes of $\pi\pi$ and $\pi$K elastic scattering in the energy
region of order of $1\,{\rm GeV}$.

Unlike other available theoretical approaches to the mechanism of
$\Lambda$K$^+$ and $\Sigma^0$K$^+$ production in pp collisions
[10,11,16] our mechanism does not demand the inclusion of exchanges of
all mesons heavier than the $\pi^0$--meson and baryon resonances
$N(1650)$, $N(1710)$ and so on.  One can show that the summary
contribution of the one--meson exchanges of $\eta(550)$, $\rho(770)$
and $\omega(780)$ meson and the scalar isoscalar meson $\sigma(700)$ [17--21]
is of order of 10$\%$ relative to the one--pion exchange.  In
fact, the estimate of the summary contribution of the $\eta(550)$,
$\sigma(700)$, $\rho(770)$ and $\omega(780)$ meson exchanges relative
to the one--pion exchange reads
\begin{eqnarray*}
&&\frac{1}{3}\, \Bigg(\frac{D - 3F}{D + F}\Bigg)^2\,
\frac{\displaystyle M^2_{\pi} + 2M_{\rm p}(\sqrt{M^2_{\rm p} + p^2_0}
- M_{\rm p})}{\displaystyle M^2_{\eta} + 2M_{\rm p}(\sqrt{M^2_{\rm p}
+ p^2_0} - M_{\rm p})}- \frac{1}{g^2_{\rm A}}\,
\frac{\displaystyle M^2_{\pi} + 2M_{\rm p}(\sqrt{M^2_{\rm p} + p^2_0}
- M_{\rm p})}{\displaystyle M^2_{\sigma} + 2M_{\rm p}(\sqrt{M^2_{\rm p}
+ p^2_0} - M_{\rm p})}\nonumber\\
&& + 2\,\frac{g^2_{\rho}}{g^2_{\rm \pi NN}}\,
\frac{\displaystyle M^2_{\pi} + 2M_{\rm p}(\sqrt{M^2_{\rm p} + p^2_0}
- M_{\rm p})}{\displaystyle M^2_{\rho} + 2M_{\rm p}(\sqrt{M^2_{\rm p}
+ p^2_0} - M_{\rm p})} = 0.05 - 0.36 +  0.22 = - 0.09 (9\%),
\end{eqnarray*}
where we have assumed that the $\eta(550)$ is the eighth component of
the octet of pseudoscalar mesons and the coupling constant of the
${\rm \sigma NN}$ interaction is equal to $g_{\rm \sigma NN} = g_{\rm
\pi NN}/g_{\rm A}$ [17] with $g_{\rm A} = 1.267$, the axial--vector
coupling constant [6]. Then, we have set $g_{\rho} = 6.047$ is the
$\rho \pi \pi$ coupling constant [6]. The value 9$\%$ can be reduced
by including the contribution of the pseudoscalar $\eta'(958)$
meson. This confirms that with a good accuracy the one--pion exchange
dominates in pp reactions for $\Lambda$K$^+$ and $\Sigma^0$K$^+$
production at thresholds of the final p$\Lambda$K$^+$ and
p$\Sigma^0$K$^+$ states.

We do not take into account the contributions of baryon resonances
$N(1650)$, $N(1710)$ and so on [16]. Nevertheless, the obtained
agreement with the experimental data allows us to think that
effectively the contributions of baryon resonances can be partly
reproduced by the amplitude of the pp interaction in the initial
state.

In our approach the daugther proton and the $\Lambda$ hyperon as well
as the daugther proton and the $\Sigma^0$ hyperon are in the
spin--singlet state. This implies that the direction of the spin of
the $\Lambda$ and $\Sigma^0$ hyperons is strictly opposite to the
direction of the spin of the daugther proton. Thereby, according to
our approach by measuring a polarization of the daughter proton one
measures unambiguously a polarization of the $\Lambda$ and $\Sigma^0$
hyperons. Of course, this is true only for the excess of energies very
close to thresholds of the final states p$\Lambda$K$^+$ and
p$\Sigma^0$K$^+$. For the excess of energies when the contribution of
the spin--triplet state of the p$\Lambda$ and p$\Sigma^0$ system
becomes perceptible the polarizations of the $\Lambda$ and $\Sigma^0$
hyperons are not so strictly determined. We are planning to carry out
the analyse of the polarization properties of the $\Lambda$ and
$\Sigma^0$ hyperons by taking into account the contribution of the
spin--triplet states of the p$\Lambda$ and p$\Sigma^0$ systems in our
forthcoming publications\,\footnote{The most complete phenomenological
analysis of the reactions p + p $\to$ p + $\Lambda$ + K$^+$ and p + p
$\to$ p + $\Sigma^0$ + K$^+$ with polarized colliding protons near
thresholds of the final states p$\Lambda$K$^+$ and p$\Sigma^0$K$^+$,
respectively, has been carried out by Rekalo {\it et al.} [22].}.

Recent measurements of the polarization of the $\Lambda$ hyperon in
the reaction p + p $\to$ p + $\Lambda$ + K$^+$ at the excess of energy
$\varepsilon = 431\,{\rm MeV}$ [23] evidence an advantage of the
K$^+$--meson exchange mechanism [24] with respect to the one--pion
exchange one. In our approach being valid for the excess of energies
much less than $\varepsilon = 431\,{\rm MeV}$ the contribution of the
K$^+$--meson exchange makes up about 0.1$\%$ in comparison with the
exchange by the $\pi^0$ meson.

\section*{Acknowledgement}

\hspace{0.2in} The work is supported in part by the Scientific and
Technical Programme of Ministry of Education of Russian Federation for
Fundamental Researches in Universities of Russia.

\newpage

\noindent Table 1. Cross sections for the reaction p + p $\to$ p +
$\Lambda$ + K$^+$ for the excess of energy ranging over the region
$0.68\,{\rm MeV} \le \varepsilon \le 6.68\,{\rm MeV}$. The
experimental data are taken from Ref.[2].

\vspace{0.2in}

\begin{tabular}{|c|c|c|c|c|c| }\hline
\cline{2-6} $\varepsilon$ & $\Omega_{\rm p\Lambda
K^+}(\varepsilon)$ & $\sigma_{\rm p\Lambda K^+}(\varepsilon)$ & $\sigma_{\rm
p\Lambda K^+}(\varepsilon)/\varepsilon^2$ & $\sigma_{\rm p\Lambda K^+}
(\varepsilon)_{\exp}$ &
$\sigma_{\rm p\Lambda K^+}(\varepsilon)_{\exp}/\varepsilon^2$ \\
       (MeV)&       &(nb)  & (nb/MeV$^2$)  &  (nb)       &(nb/MeV$^2$)\\ 
\hline 0.68 &0.516  & $1.8\pm 0.1$&  $4.0\pm 0.1$ &$2.1\pm 0.2$ & 4.54\\ 
       1.68 &0.605  & $13.2\pm 0.3$& $4.7\pm 0.1$ &$13.4\pm0.7$ & 4.75\\ 
       2.68 &0.616  & $34.1\pm 0.7$& $4.8\pm 0.1$ &$36.6\pm2.6$ & 5.10 \\ 
       3.68 &0.609  & $63.6\pm 1.2$& $4.7\pm 0.1$ &$63.0\pm3.1$ & 4.65 \\ 
       4.68 &0.594  & $100.3\pm 1.9$&$4.6\pm 0.1$&$92.2\pm6.5$ & 4.21 \\ 
       5.68 &0.577  & $143.5\pm 2.8$&$4.5\pm 0.1$&$135\pm11$   & 4.18 \\ 
       6.68 &0.560  & $192.6\pm 3.7$&$4.3\pm 0.1$&$164\pm10$ & 3.68 \\ 
\hline      &       &       & $4.5\pm 0.1$       &       & $4.4\pm 0.7$ \\ 
\hline
\end{tabular}\\

\noindent Table 2. Cross section for the reaction p + p $\to$ p +
$\Sigma^0$ + K$^+$ for the excess of energy ranging over the region
$0.68\,{\rm MeV} \le \varepsilon \le 6.68\,{\rm MeV}$. The
experimental data are taken from Ref.[4].

\vspace{0.2in}

\begin{tabular}{|c|c|c|c|c|c| }\hline
\cline{2-6} $\varepsilon$ & $\Omega_{\rm p\Sigma^0 K^+}(\varepsilon)$ &
$\sigma_{\rm p\Sigma^0 K^+}(\varepsilon)$ & $\sigma_{\rm
p\Sigma^0 K^+}(\varepsilon)/\varepsilon^2$ & 
$\sigma_{\rm p\Sigma^0 K^+}(\varepsilon)_{\exp}$ &
$\sigma_{\rm p\Sigma^0 K^+}(\varepsilon)_{\exp}/\varepsilon^2$ \\
      (MeV) &        &  (nb) & (nb/MeV$^2$)&(nb)&(nb/MeV$^2$)\\
\hline 0.68 &0.515  & $0.11\pm 0.01$& $0.23\pm 0.03$ &$0.14\pm 0.06$&$0.29\pm 0.14$ \\ 
       1.68 &0.603  & $0.80\pm 0.10$& $0.27\pm 0.03$ &$0.73\pm 0.34$&$0.26\pm 0.12$ \\ 
       2.68 &0.613  & $2.00\pm 0.25$& $0.28\pm 0.03$ &$1.67\pm 0.77$&$0.23\pm 0.11$\\ 
       3.68 &0.605  & $3.71\pm 0.46$& $0.28\pm 0.03$ &$2.87\pm 1.32$&$0.21\pm 0.10$\\ 
       4.68 &0.590  & $5.85\pm 0.72$& $0.27\pm 0.03$  &$4.26\pm 1.97$&$0.20\pm 0.09$\\ 
       5.68 &0.572  & $8.36\pm 1.03$& $0.26\pm 0.03$ &$5.83\pm 2.69$&$0.18\pm 0.08$\\ 
       6.68 &0.555  & $11.29\pm 1.38$&$0.25\pm 0.03$ &$7.53\pm 3.47$&$0.17\pm 0.08$ \\ 
\hline    &         &       & $0.26\pm 0.03$  &    & $0.22\pm 0.11$\\ \hline
\end{tabular}\\

\newpage

\section*{Figure caption}

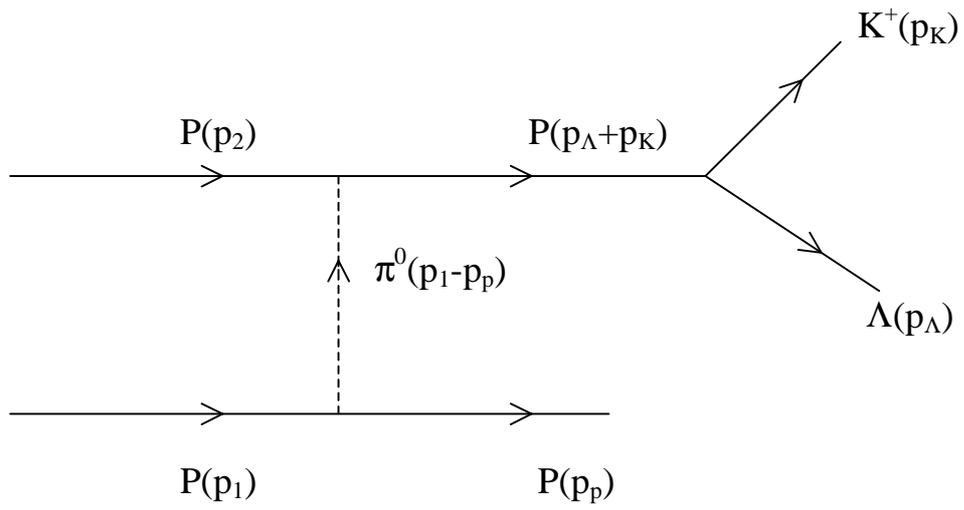
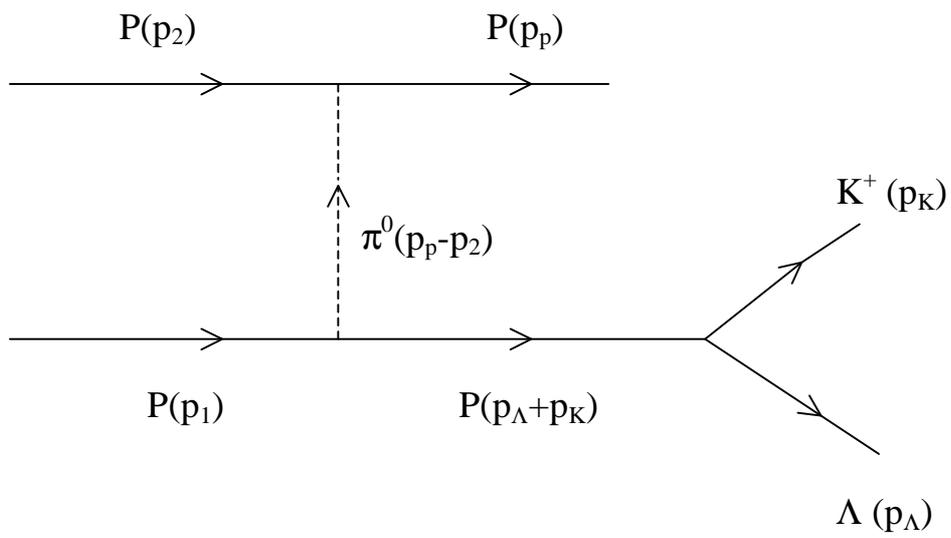
Fig1. The one--pion exchange diagrams describing the effective
Lagrangian of the low--energy transition p + p $\to$ p + $\Lambda$ + K$^+$.

\end{document}